\documentclass[conference]{IEEEtran}
\IEEEoverridecommandlockouts

\usepackage{amsmath,amssymb,amsfonts}
\usepackage{algorithmic}
\usepackage{graphicx}
\usepackage{subcaption}
\usepackage[colorlinks=true, linkcolor=red, citecolor=red, urlcolor=blue]{hyperref}
\usepackage{textcomp}
\usepackage{xcolor}
\usepackage[numbers,sort&compress]{natbib}
\usepackage{fancyhdr}

\def\BibTeX{{\rm B\kern-.05em{\sc i\kern-.025em b}\kern-.08em
    T\kern-.1667em\lower.7ex\hbox{E}\kern-.125emX}}
\begin{document}

\title{In-place Switch: Reprogramming based SLC Cache Design for Hybrid 3D SSDs}

\author{
\IEEEauthorblockN{Xufeng Yang}
\IEEEauthorblockA{\textit{Xiamen university},
Xiamen, China \\
xufengyang@stu.xmu.edu.cn}
\and
\IEEEauthorblockN{Jiancong Zheng}
\IEEEauthorblockA{\textit{Xiamen university},
Xiamen, China \\
zhengjc@stu.xmu.edu.cn}
\and
\IEEEauthorblockN{Congming Gao*}
\IEEEauthorblockA{\textit{Xiamen university},
Xiamen, China \\
gaocm92@gmail.com}
}

\maketitle
\vspace*{-16pt}
\begin{abstract}
Recently, 3D SSDs have been widely adopted in PCs, data centers, and cloud storage systems. High-bit-density cells, such as Triple-Level Cell (TLC), are utilized within 3D SSDs to increase capacity. However, due to the inferior performance of TLC, a portion of TLCs is configured to operate as Single-Level Cells (SLC) to provide high performance, with host data initially directed to the SLCs. In SLC/TLC hybrid 3D SSDs, a portion of the TLC space is designated as an SLC cache to achieve high SSD performance by writing host data at the SLC speed. Given the limited size of the SLC cache, block reclamation is necessary to free up the SLC cache during idle periods. However, our preliminary studies indicate that the SLC cache can lead to a performance cliff if filled rapidly and cause significant write amplification when data migration occurs during idle times.

In this work, we propose leveraging a reprogram operation to address these challenges. Specifically, when the SLC cache is full or during idle periods, a reprogram operation is performed to switch used SLC pages to TLC pages in place (termed In-place Switch, IPS). Subsequently, other free TLC space is allocated as the new SLC cache. IPS can continuously provide sufficient SLC cache within SSDs, significantly improving write performance and reducing write amplification. Experimental results demonstrate that IPS can reduce write latency and write amplification by up to 0.75 times and 0.53 times, respectively, compared to state-of-the-art SLC cache technologies.
\end{abstract}

\begin{IEEEkeywords}
 Hybrid 3D SSDs, SLC Cache, TLC, Reprogram.
\end{IEEEkeywords}

\begin{figure*}[ht]
  \centering
    \begin{minipage}[b]{0.98\textwidth}
      \includegraphics[width=0.98\columnwidth]{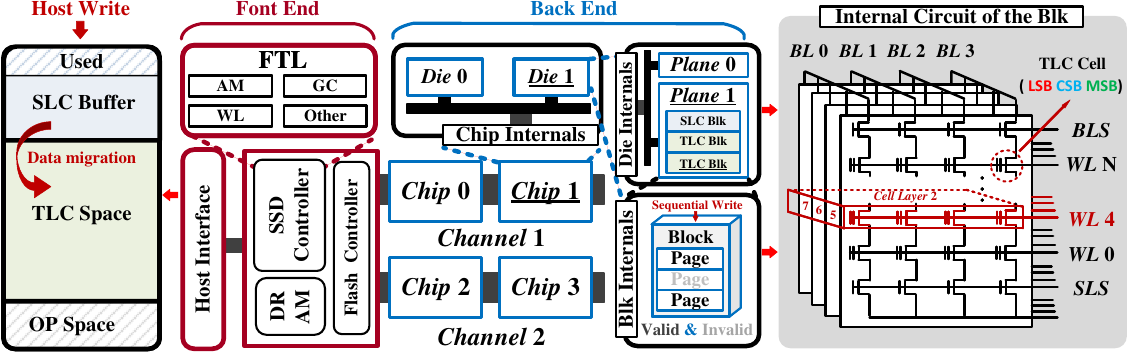}
    \end{minipage}
\caption{The architecture of 3D hybrid SSD.}
\label{fig:ssd_arch}
    \vspace{-0.15in}
\end{figure*}

\vspace{-9pt}
\section{\textbf{Introduction}}

Recently, NAND flash memory-based SSDs have been widely adopted in personal computers, data centers, and cloud storage systems due to their high performance, large capacity, and low power consumption.
To enhance the capacity of SSDs, high-bit-density cells, such as triple-level cells (TLC) or quad-level cells (QLC), are utilized to store more bits per cell. Additionally, various three-dimensional (3D) architectures of NAND flash memory, in conjunction with high bit-density cells (with TLC as the focus in this discussion),  have been launched into the market to overcome the scaling limit of planar NAND flash arrays \cite{goda2021recent,du2014overview, Lee2021ArchitectureAP}. However, there is no such thing as a free lunch; while 3D TLC SSDs significantly increase storage density, they come with poorer performance. Compared to single-level cell (SLC) or multi-level cell (MLC) based SSDs, 3D TLC SSDs require more time to decode (read) or encode (write) data.  

As an alternative solution, SLC caches are integrated into commercial 3D TLC SSDs to mitigate this performance issue \cite{dynamicwritemicron,inteldy,samsungturbo}. During host writes, data is first written to a limited-capacity SLC cache, leveraging the higher performance of SLCs to deliver better overall SSD performance.  SLC cache is cleared during idle time to sustainably provide enough space for following writes. However, two key challenges arise from the characteristics of SLC caches: 
1) The limited SLC cache can quickly become saturated under sustained write workloads. Due to the low bit density of SLC, hybrid 3D SSDs typically allocate only a small portion of TLC space as an SLC cache, preserving the capacity advantage of hybrid 3D SSDs. Once saturated, sustained writes are redirected to the TLC space and are processed at TLC performance levels, {\em \textbf{leading to a performance cliff}};
2) The reclamation process for the SLC cache results in a large amount of additional writes. The SLC cache is cleared by migrating data from used SLC pages to available TLC space and erasing used blocks during idle time. Since NAND flash memory is a wear-sensitive storage medium, this additional data migration leads to {\em \textbf{noticeable write amplification increase}} \cite{yoo2020reinforcement, shin2012new, hu2009write}.

To overcome the above mentioned problems, we present \underline{\textbf{In-place Switch (IPS)}}, which adopts reprogram operations to reuse the SLC cache when it is full. Firstly, IPS continually attempts to allocate a new SLC cache to prevent performance cliffs and ensure sustained high write performance. Additionally, IPS reprograms used SLC pages into TLC pages for switching SLC pages to TLC pages in place, reducing the data migration associated with SLC cache reclamation. This process maintains the capacity advantage of hybrid SSDs. IPS adheres to the limitations of the reprogram operation in 3D TLC SSDs, and the feasibility of the reprogram operation in 3D TLC SSDs has been validated\cite{gao2019constructing}. The results reveal that IPS can effectively improve the performance and reduce the write amplification of hybrid 3D SSDs.
On average, write latency is reduced by 0.75 times, and write amplification is reduced by 0.53 times at most.
We summarized our contributions as follows:
\vspace{-1pt}
\begin{itemize}
\item
We analyze the performance and write amplification of SLC/TLC hybrid 3D SSDs, and then point out the potential pitfalls of hybrid 3D SSDs;
\item
We propose an In-place Switch based SLC cache design for hybrid 3D SSDs to overcome the disadvantages of conventional hybrid 3D SSDs;
\item
We evaluate IPS in a modified hybrid 3D SSDs-based simulator \cite{hu2012exploring}.
The experimental results show that IPS effectively improves performance and reduces the write amplification of hybrid 3D SSDs.
\end{itemize}

\section{\textbf{Background and Related Work}}

\subsection{\textbf{3D NAND Flash Memory based SSD}}

Figure \ref{fig:ssd_arch} illustrates the architecture of a hybrid 3D SSD that integrates SLC and TLC spaces within a single storage device. The SSD architecture comprises two parts: 1) The front end consists of the control and communication units. The SSD controller handles read and write requests from the host interface and executes Flash Translation Layer (FTL) functions such as Address Mapping (AM), Garbage Collection (GC), and Wear Leveling (WL). The flash controller serves as the interface between the front end and the back end; 2) The back end consists of the data storage units that provide actual storage capacity. To boost SSD performance, the internal architecture employs four levels of parallelism: from channel to chip, die, and plane\cite{hu2012exploring}. Each plane contains multiple blocks, which can only be written sequentially.

The primary distinction between 3D SSDs and 2D SSDs lies in the architecture of the flash block.  As shown on the right of Figure \ref{fig:ssd_arch}, 3D SSDs employ a vertical block architecture.  To transition from 2D to 3D, 3D SSDs introduce layers, with each layer consisting of multiple word lines on a horizontal plane, and each word line comprising multiple flash cells. TLC flash provides three types of bits for each cell, Least-Significant-Bit (LSB), Center-Significant-Bit (CSB), and Most-Significant-Bit (MSB). Traditionally, program operations shift the voltage state from "erased" to a specific higher voltage level, representing three-bit values in TLC (LSB, CSB, MSB). Using three registers, one-shot programming enables the simultaneous writing of three pages to a single word line, a process known as "one-shot programming" \cite{ho2018achieving}. SSDs employ an out-of-place update strategy for program operations, where updates are written to free pages while the original pages are invalidated. A cell can only be reprogrammed after it has been erased.

\subsection{\textbf{Reprogram Operation in 3D TLC SSDs}}
\begin{figure}[!htb]
    \vspace{-0.1in}
  \centering
    \begin{minipage}[b]{0.45\textwidth}
      \includegraphics[width=0.98\columnwidth]{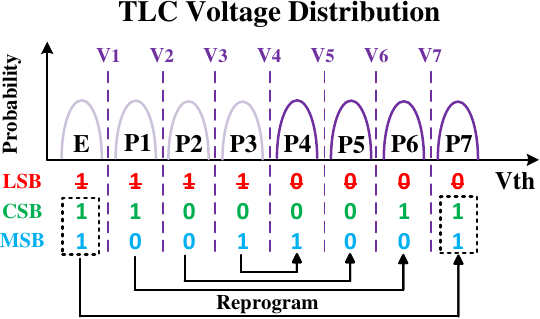}
    \end{minipage}
\caption{An example of how reprogram operation accelerate the read process of flash cell.}
\label{fig:bg_repro}
    \vspace{-0.2in}
\end{figure}

Previous works \cite{long2022adar, choi2018invalid, chang2016realizing, choi2017exploiting, gao2019constructing, wu2018flash} have developed reprogram operations for both 2D and 3D flash memory. Reprogram operation refers to the process of reprogram flash cells that are at a certain voltage state to a higher voltage state before they are erased, thereby achieving in-place updates to reduce cell wear or improving the reading process. 
As shown in Figure \ref{fig:bg_repro}, this process can accelerate reads in TLC SSDs \cite{choi2018invalid}. The x-axis represents the voltage threshold, and the y-axis indicates the probability of each state. By charging flash memory cells with different amounts of electrons, the entire voltage range can be divided into eight states (E, P1$\sim$P7). 
If LSB bits are invalidated, the eight states of CSB and MSB bits can be grouped into four pairs (E-P7, P1-P6, P2-P5, P3-P4) due to identical data. Reprogramming the voltage thresholds in the first four states (E$\sim$P3) to the latter four states (P4$\sim$P7) can accelerate the read process \cite{pan2012quasi}.

\begin{figure*}[ht]
  \vspace{-0.15in}
  \subfloat[Crucial P1.]{
    \label{fig:bw-c}
    \begin{minipage}[b]{0.33\textwidth}
      \includegraphics[width=0.96\columnwidth]{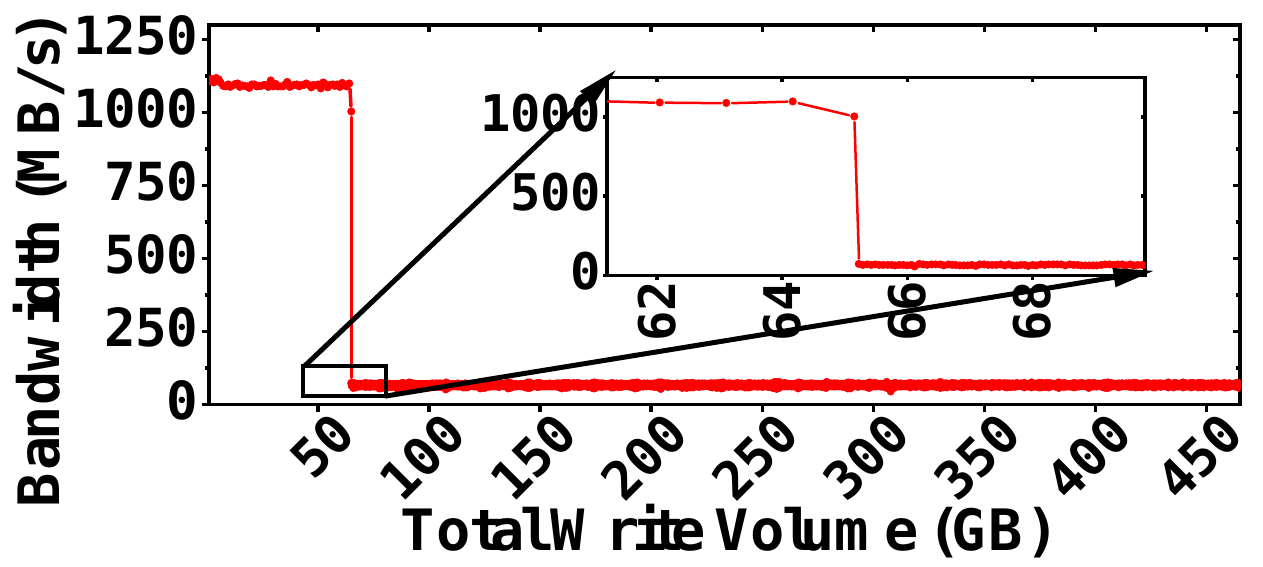}
      \vspace{-0.06in}
    \end{minipage}}
  \vspace{-0.14in}
  \hspace{-0.15in}
  \subfloat[Intel 660p.]{
    \label{fig:bw-i}
    \begin{minipage}[b]{0.33\textwidth}
      \includegraphics[width=0.96\columnwidth]{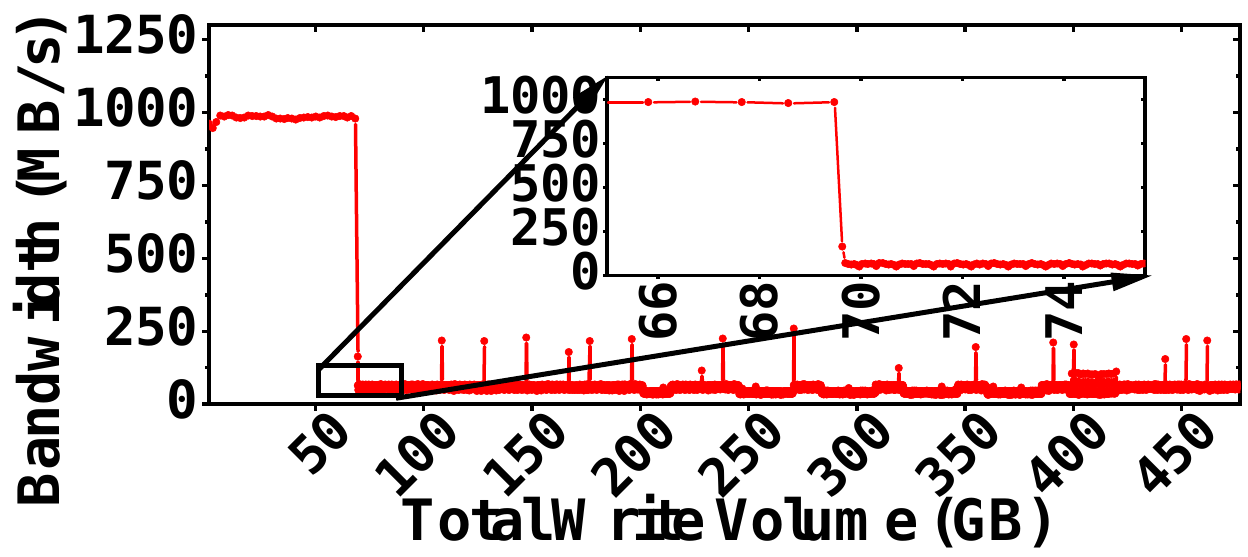}
      \vspace{-0.06in}
    \end{minipage}}
  \vspace{-0.14in}
  \hspace{-0.15in}
  \subfloat[Samsung 970 EVO Plus.]{
    \label{fig:bw-s}
    \begin{minipage}[b]{0.33\textwidth}
      \includegraphics[width=0.96\columnwidth]{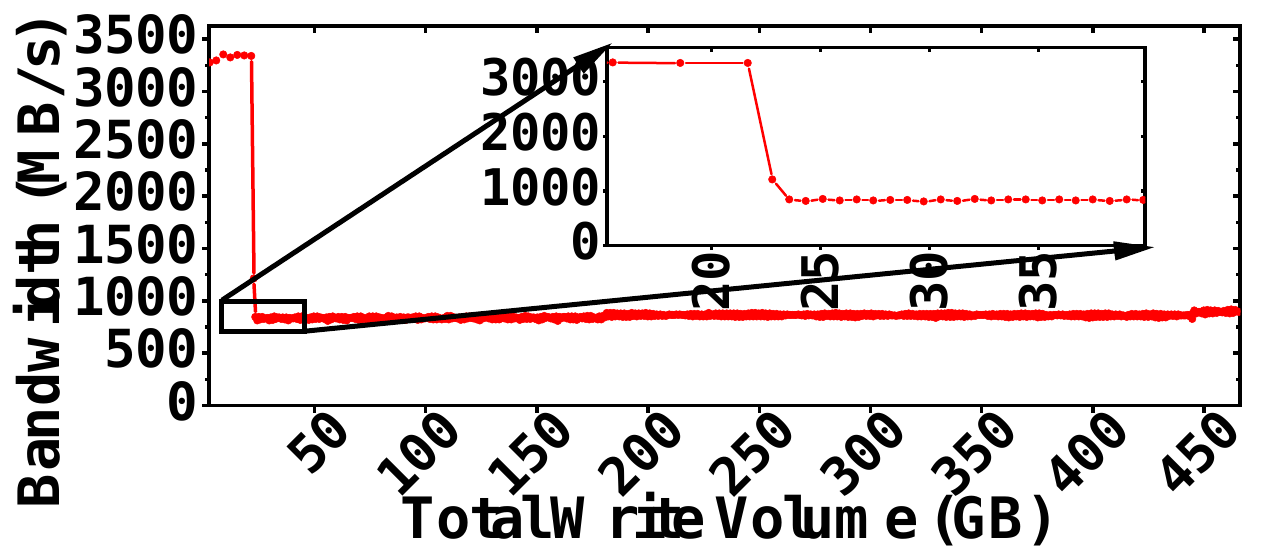}
      \vspace{-0.06in}
    \end{minipage}}
    \vspace{0.2in}
\caption{The bandwidths of SSDs with sustained sequential writes.}
\label{fig:bw}
\end{figure*}

\begin{figure*}[ht]
  \vspace{-0.1in}
  \subfloat[Crucial P1.]{
    \label{fig:dw-c}
    \begin{minipage}[b]{0.334\textwidth}
      \includegraphics[width=0.96\columnwidth]{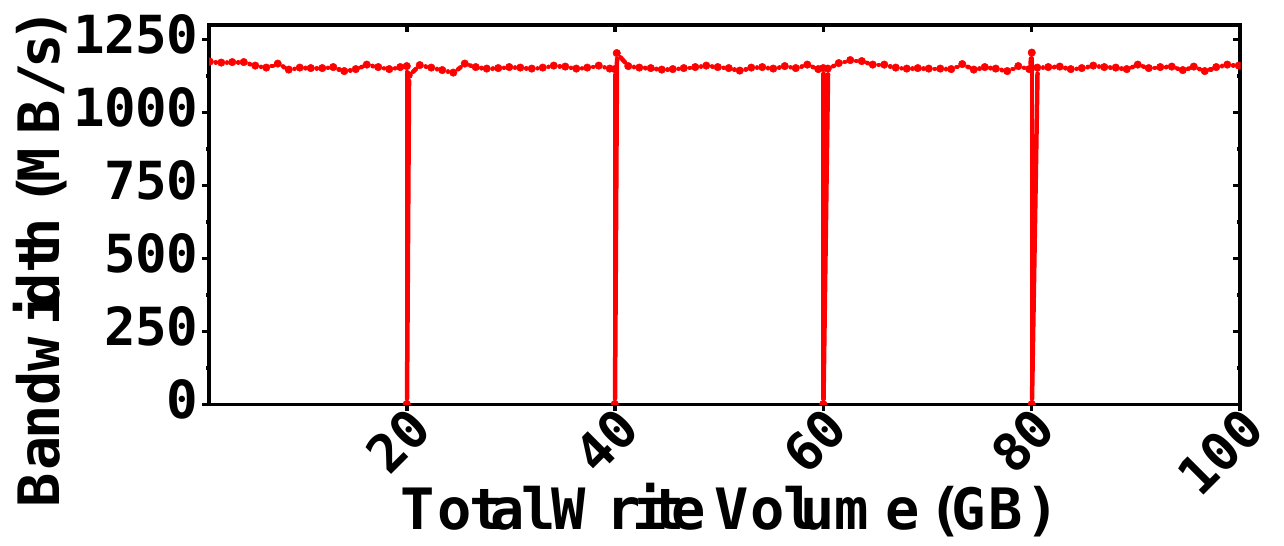}
      \vspace{-0.06in}
    \end{minipage}}
    \vspace{-0.1in}
    \hspace{-0.17in}
  \subfloat[Intel 660p.]{
    \label{fig:dw-i}
    \begin{minipage}[b]{0.337\textwidth}
      \includegraphics[width=0.96\columnwidth]{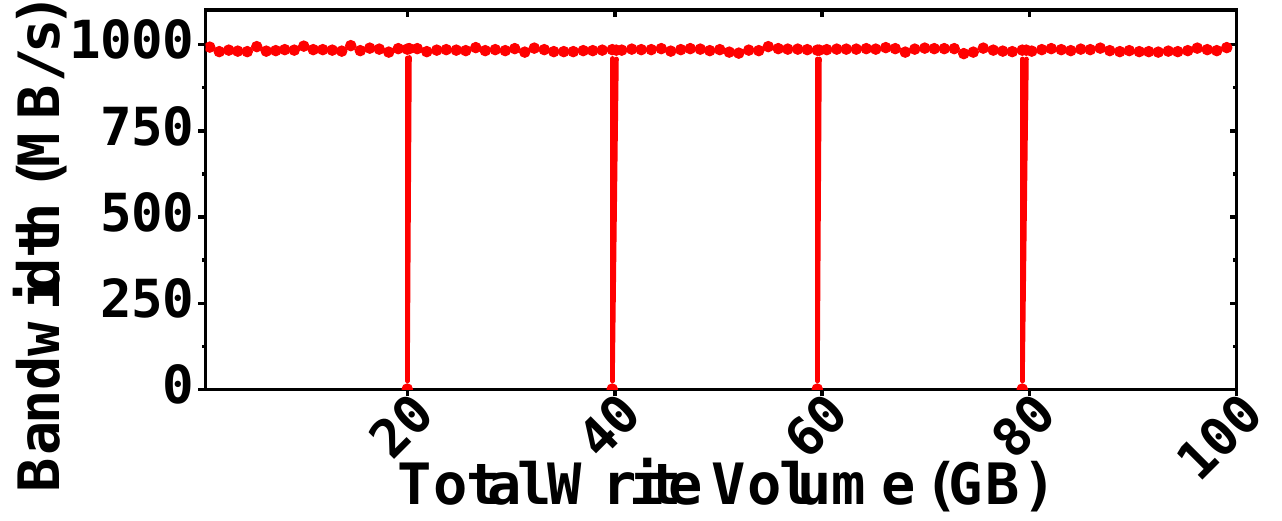}
      \vspace{-0.06in}
    \end{minipage}}
    \hspace{-0.17in}
  \subfloat[Samsung 970 EVO Plus.]{
    \label{fig:dw-s}
    \begin{minipage}[b]{0.335\textwidth}
      \includegraphics[width=0.96\columnwidth]{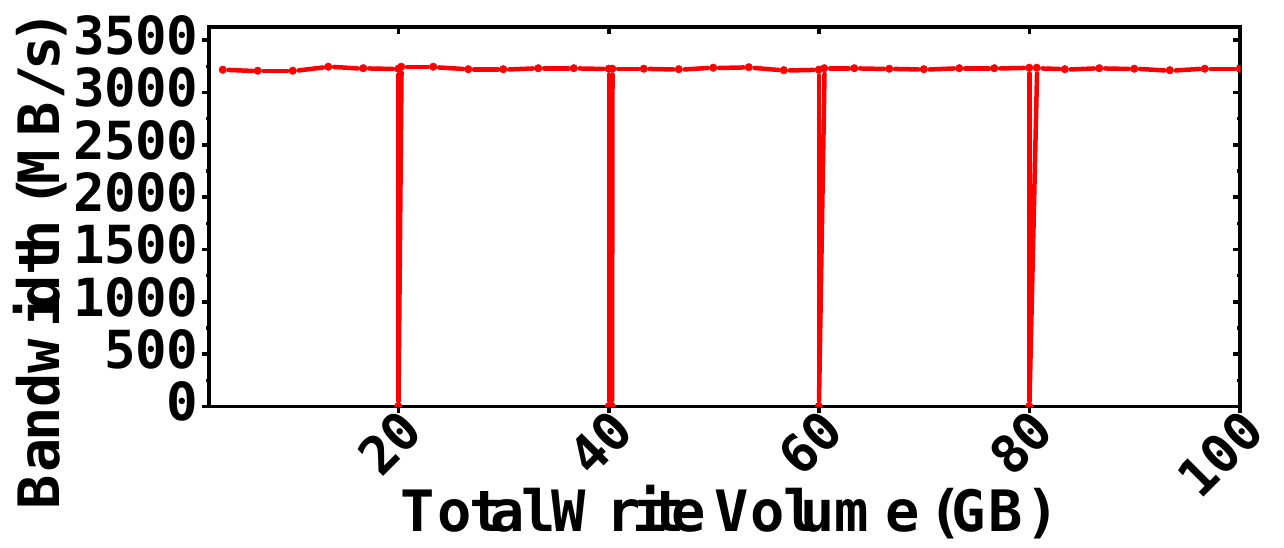}
      \vspace{-0.06in}
    \end{minipage}}
    \vspace{0.02in}
\caption{The bandwidths of SSDs with periodical sequential writes.}
\vspace{-0.2in}
\label{fig:dw}
\end{figure*}

However, two main challenges affect the reliability of reprogram operations: cell-to-cell interference \cite{cai2017vulnerabilities} and background pattern dependency \cite{lee20167}. The former has a relatively minor impact, while the latter plays a more significant role in reducing reliability. Therefore, to maintain the reliability of flash memory, reprogram operations must be performed with caution. According to the verification of reprogram operation presented in \cite{gao2019constructing}, there are two restrictions for performing the technology: First, 3D TLC only can be randomly reprogrammed within two layers in 3D TLC SSDs without loss of reliability; Second, each TLC can be reprogrammed four times at most. In this work, these restrictions are strictly satisfied when reprogram operation is considered.

\subsection{\textbf{SLC Cache in Hybrid 3D SSDs}}
Modern commodity hybrid 3D SSDs combine SLC and TLC \cite{crucialp1,shin2012new}, where SLC is used as the cache to store user data for boosting the performance of hybrid SSDs. As shown in the left part of Figure \ref{fig:ssd_arch}, during host writes, data is first written to the limited-capacity SLC cache. This approach leverages the higher performance of SLCs, resulting in improved overall SSD performance. Meanwhile, the TLC space ensures that the 3D SSDs maintain their high-density advantage, providing ample storage capacity despite the lower performance and reduced reliability compared to the SLC cache. When the SLC cache is full, the SSD controller performs GC, which involves migrating valid data and erasing used blocks to reclaim the SLC cache. Note that GC operations occur whenever SSD physical space is insufficient, not just when the SLC cache is full.

Since the SLC cache is engineered by only storing one bit in the original TLC, the size of the SLC cache should be carefully dimensioned. Allocating more TLC space to the SLC cache enhances performance but reduces overall capacity. Conversely, allocating less TLC space to the SLC cache preserves capacity but limits performance gains from the SLC cache. Traditionally, the SLC cache is allocated from a fixed portion of TLC space, preventing excessive consumption of TLC capacity.  Typically, the size of the SLC cache in commodity SSDs ranges from several gigabytes (GBs) to over a hundred GBs \cite{dynamicwritemicron,inteldy,samsungturbo}. For example, Samsung's Turbo Write technology suggests that a 3GB SLC cache is sufficient for most daily use scenarios \cite{samsungturbo}. Other manufacturers, such as Micron, allocate larger SLC caches to accommodate data-intensive applications \cite{dynamicwritemicron}.

\subsection{Related Work}
Previous works on hybrid SSDs can be categorized into several types:
First, the size of the SLC cache is adjusted by considering the amount of data stored in the SSD or SSD's status, thus better performance is achieved without significantly losing the capacity \cite{yoo2020reinforcement,samsungturbo,dynamicwritemicron, shi2021understanding};
Second, the SLC cache can be used to accelerate time-consuming activities, such as GC, so that the performance of SSDs is boosted \cite{li2019accelerating, zhang2019spa};
Third, the SLC cache is leveraged to store hot data, thus most GCs are triggered in SLC cache as it has better endurance and lower read and write latencies \cite{im2010comboftl,liu2017workload,li2022latency};
Fourth, to balance the wear rates of SLC and TLC, Jimenez {\em et al.} \cite{jimenez2015libra} proposes to distribute writes to the SLC cache or TLC space according to flash cells' wear rate.

However, above mentioned works still need block reclamation to empty the SLC cache, which does harm to performance and lifetime of SSDs from the perspective of write latency and write amplification.
Therefore, in this work, we attempt to alleviate block reclamation's impact via reprogram operation.

\section{\textbf{Motivation and Problem Statement}}\label{sec:moti}
To analyze the characteristics of SLC cache in hybrid SSD, a real 3D hybrid SSD with 500GB \cite{crucialp1} is evaluated with two scenarios: bursty access and daily use. 
For bursty access, the SSD is accessed by sustained sequential writes without idle time and its bandwidth is measured, as shown in Figure \ref{fig:bw}.
For daily use, five distinct sequential write streams are executed in order, each of which writes 20GB data into the SSD. 
A free period of 10 minutes is introduced between consecutive write streams to allow the device to idle, which is large enough as Turbo Write technology adopted by Samsung sets the idle time to smaller than 1 minute \cite{samsungturbo}.
The measured bandwidth for the daily use scenario is presented in Figure \ref{fig:dw}.

For bursty access, as denoted in Figure \ref{fig:bw-c}, a bandwidth cliff is evident when a total 65GB data is written.
After this point, the bandwidth is significantly decreased.
Thus, two conclusions can be drawn: First, the size of the SLC cache in this SSD is approximately 65GB. Second, when the SLC cache is filled and there is no idle time to reclaim the used SLC cache, the performance is markedly reduced as subsequent writes are performed at the TLC performance level. 
However, in the daily use scenario, as illustrated in Figure \ref{fig:dw}, the bandwidth remains steady at around 1090 MB/s throughout the process, even when the total write size exceeds 65GB. This is because the SLC cache is cleared during idle periods, allowing it to be reused for subsequent writes.

\begin{figure}[!htb]
  \vspace{-0.1in}
  \subfloat[Bursty Access.]{
    \label{fig:wa-b}
    \begin{minipage}[b]{0.23\textwidth}
      \includegraphics[width=0.99\columnwidth]{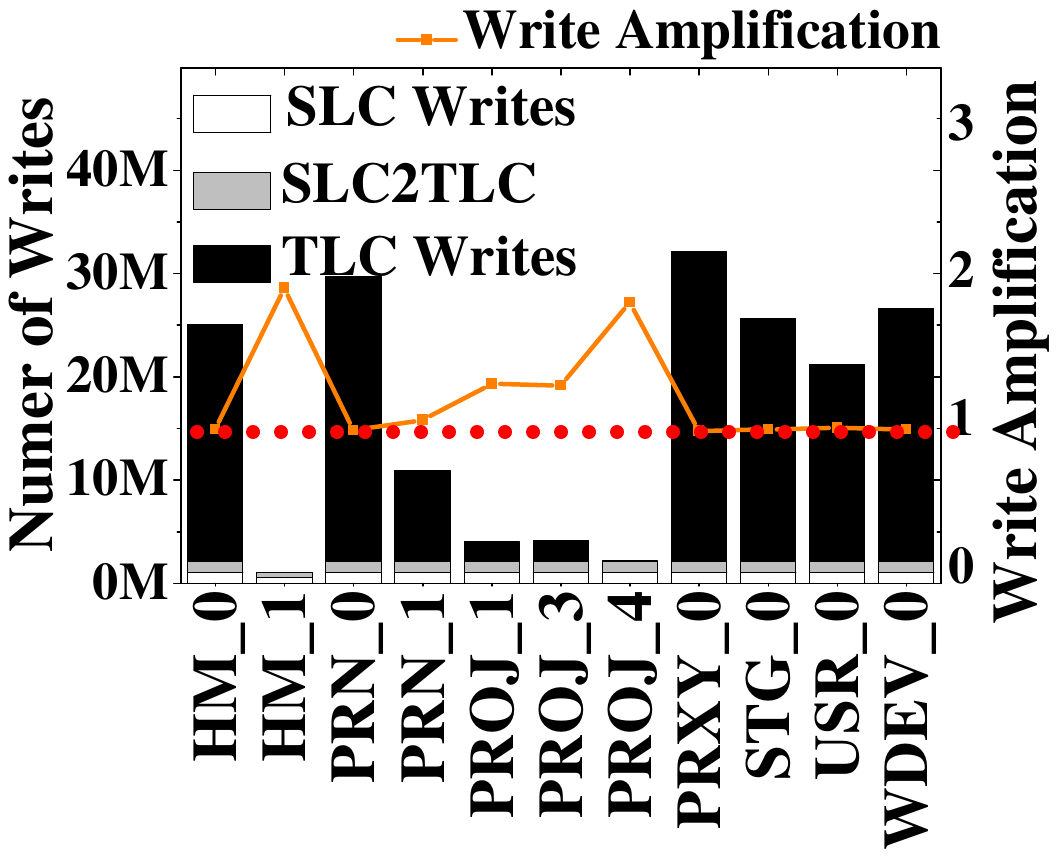}
      \vspace{-0.23in}
    \end{minipage}}
  \subfloat[Daily Use.]{
    \label{fig:wa-d}
    \begin{minipage}[b]{0.23\textwidth}
      \includegraphics[width=0.99\columnwidth]{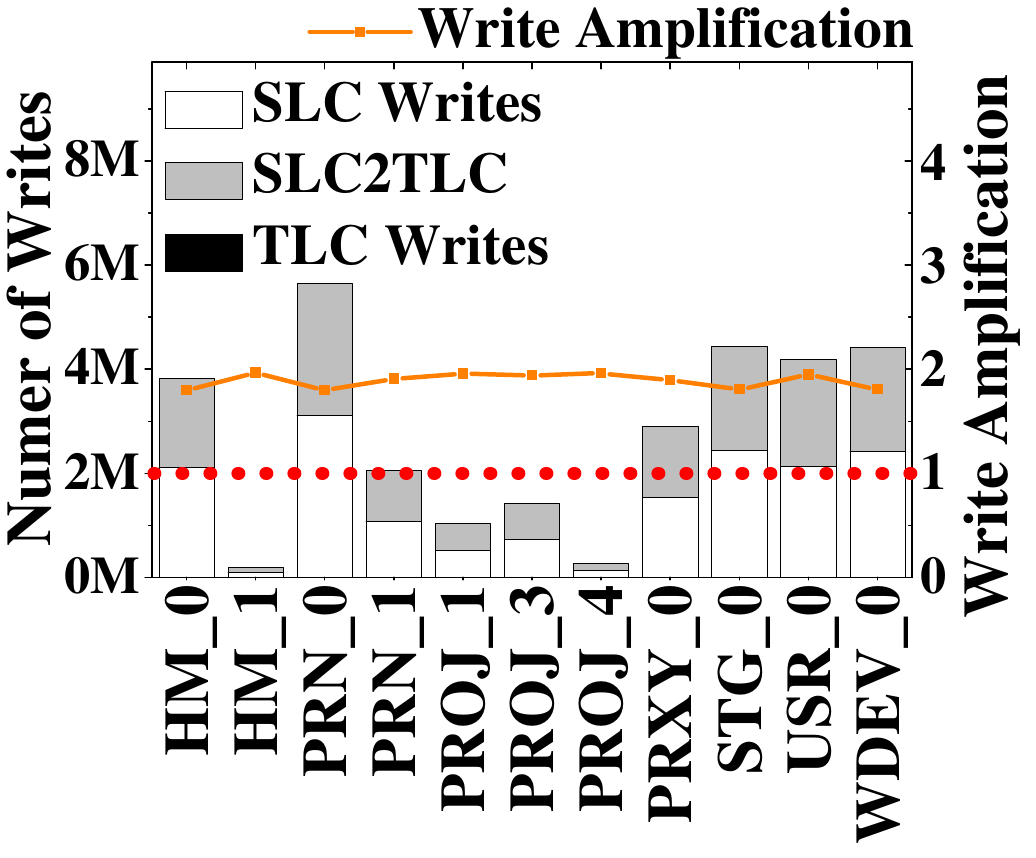}
      \vspace{-0.23in}
    \end{minipage}}
    \vspace{-0.05in}
\caption{Results of Writes Breakdown and Write Amplification.}
\vspace{-0.1in}
\label{fig:wa}
\end{figure}

However, in daily use scenarios, better performance is achieved at the penalty of frequent data migration.
To further elaborate data migration in the SLC cache, a workload-driven SSD simulator \cite{hu2012exploring} is adopted in this work, which can quantitatively analyze the number of pages migrated during idle time.
We simulate a 500GB SSD configured with a 4GB SLC cache, as the total write size of the evaluated workloads is small. To ensure a comprehensive experiment, we evaluate the simulated SSD not only under daily use scenarios with various workloads\cite{Narayanan2009} but also under bursty access scenarios.
To simulate a bursty access scenario, incoming writes should be reconstructed to fill up the SLC cache without idle time.
Thus, incoming writes of all workloads are configured as sequential writes with 32KB write size.
And then, arriving time is accelerated so that there is no idle time.
For the daily use scenario, regular workloads are executed. Additionally, at the end of each workload, all data in the SLC cache is migrated to the TLC space, and the used blocks are erased. 
The results are collected and presented in Figure \ref{fig:wa}, where all writes are categorized into three parts: writing to SLC cache (SLC Writes), data migration from SLC cache to TLC space (SLC2TLC) and directly writing to TLC space (TLC Writes).

For bursty access, as shown in Figure \ref{fig:wa-b}, TLC writes, which do not cause write amplification, dominate the majority of workloads (9 out of 11). 
As a result, write amplifications of most workloads only can be slightly reduced as SLC2TLC writes only account for only a small portion.
On the contrary, for daily use, as shown in Figure \ref{fig:wa-d}, there is a lack of TLC Writes because the SLC cache is frequently reclaimed during idle time.
Thus, write amplifications of all workloads are significantly increased, which exceeds 1.9 as the worst one reaches 1.997.

Based on the above evaluations, two potential pitfalls in hybrid SSDs arise:
First, hybrid SSDs will lead to a performance cliff in bursty access;
Second, hybrid SSDs leave a severe write amplification problem in daily use scenarios.

\section{\textbf{IPS: In-place Switch Design}}
To address these challenges, we introduce a reprogramming-based SLC cache design known as \textbf{In-place Switch (IPS)}. Initially, the fundamental concept behind IPS is explained. Following this, advanced garbage collection techniques are utilized to enhance IPS's efficiency. Additionally, we propose a design that combines advanced GC-assisted IPS with traditional SLC cache to expand the SLC cache size. Finally, we present a comprehensive discussion of the proposed approach.

\begin{figure}[ht]
    \vspace{-0.1in}
    \centering
    \begin{subfigure}[b]{0.47\textwidth}
        \includegraphics[width=0.98\columnwidth]{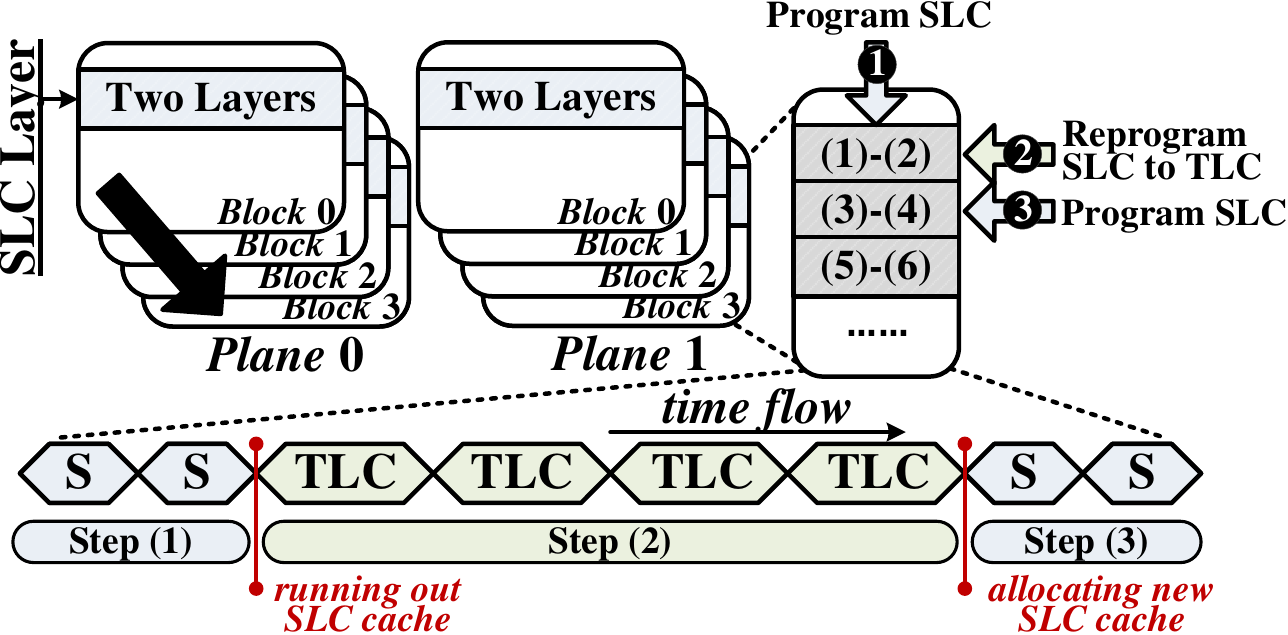}
        \caption{An example of reprogramming-based SLC cache allocation.}
        \label{fig:ov}
    \end{subfigure}
    \begin{subfigure}[b]{0.45\textwidth}
        \includegraphics[width=0.98\columnwidth]{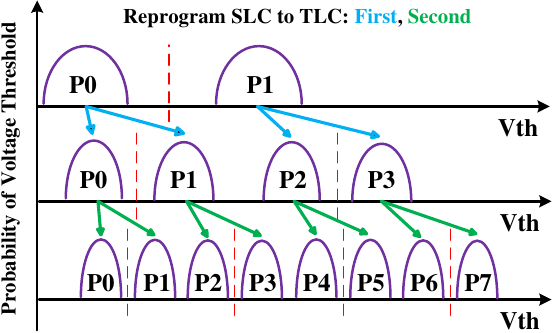}
        \caption{Voltage threshold change in reprogram operation.}
        \label{fig:d_rep}
    \end{subfigure}
    \caption{The overview of IPS design.}
    \label{fig:overview}
    \vspace{-0.15in}
\end{figure}

\subsection{\textbf{Basic Idea of IPS}}

IPS seeks to allocate a new SLC cache rather than reclaim the used SLC cache. To prevent excessive allocation of free TLC space, a new SLC cache is only allocated when the used SLC cache is fully reprogrammed, ensuring no capacity is wasted. Thus, in this design, if the SLC cache is full, host writes are utilized to reprogram SLC pages, converting them into TLC pages in place. This process not only continuously provides free SLC pages but also avoids additional data migration. Figure \ref{fig:d_rep} illustrates the changes in voltage thresholds as SLC pages are reprogrammed into TLC pages. To construct the SLC cache, only two states are encoded in each TLC cell, storing one bit per cell. IPS initially programs the flash cells into the two lower voltage thresholds (compared to traditional SLC) to represent the two SLC states. If reprogram operation is needed to convert SLC back to TLC in place, the original data is first read and then rewritten with new data. Reprogramming the lower two voltage thresholds used for the SLC cache into the higher eight voltage thresholds required for TLC involves two reprogram operations.

However, considering the restrictions of reprogram operation \cite{gao2019constructing}, the SLC cache is allocated at the granularity of two layers per block.
Initially, the first two layers of all blocks are allocated as SLC cache to provide basic SLC cache capacity.
As shown in Figure \ref{fig:ov}, the first two layers (termed SLC layer) of all blocks in different planes (denoted as Plane 0 and Plane 1) are allocated and grouped as SLC cache.
At the beginning, within each plane, all blocks containing the SLC layer are programmed sequentially to store data in SLC pages (\textbf{Step 1}).
After all SLC layers in a plane have been exhausted, reprogram operations are performed in \textbf{Step 2}, requiring four operations to convert two SLC pages back to TLC pages.  Once an SLC layer has been fully reprogrammed, the next two layers in the same block are allocated as new SLC layers.
Consequently, subsequent writes can be processed at the SLC performance level again (\textbf{Step 3}).

For bursty access, IPS enhances performance over traditional SLC caches, as host writes benefit from the new SLC cache. For daily use, write amplification is reduced by substituting traditional idle-time data migration with reprogram operations, avoiding additional writes to the SSDs. However, reprogram operations are performed at the TLC performance level, leading to a decrease in performance. To address this, we propose leveraging advanced garbage collection techniques \cite{jung2012taking} to regain the high-performance benefits of the SLC cache while preserving the reduction in write amplification achieved by IPS.

\subsection{\textbf{Advanced GC assisted IPS}}
Advanced GC (AGC) aims to divide GC into several atomic processes including multiple valid page migrations and erase operations.
By moving these processes from runtime to idle time, AGC can hide its time cost and improve the performance of SSDs.
In this work, IPS is designed with considering AGC (termed IPS/agc).
IPS/agc leverages valid page migration of AGC to reprogram used SLC pages during idle time.
In detail, instead of migrating data from the SLC cache to free TLC space, valid pages of AGC are read and reprogrammed to used SLC pages so that a new SLC layer can be allocated during idle time.
IPS/agc can outperform traditional SLC cache due to three main reasons:
First, no additional data migration occurs so that the write amplification IPS achieves can be maintained;
Second, used SLC pages can be reprogrammed more frequently during idle time so that more new SLC cache can be allocated to store host writes in runtime;
Third, valid page migration from AGC blocks can be flexibly interrupted to barely delay host writes.

\begin{figure}[ht]
    \vspace{-0.1in}
  \centering
    \begin{minipage}[b]{0.49\textwidth}
      \includegraphics[width=0.98\columnwidth]{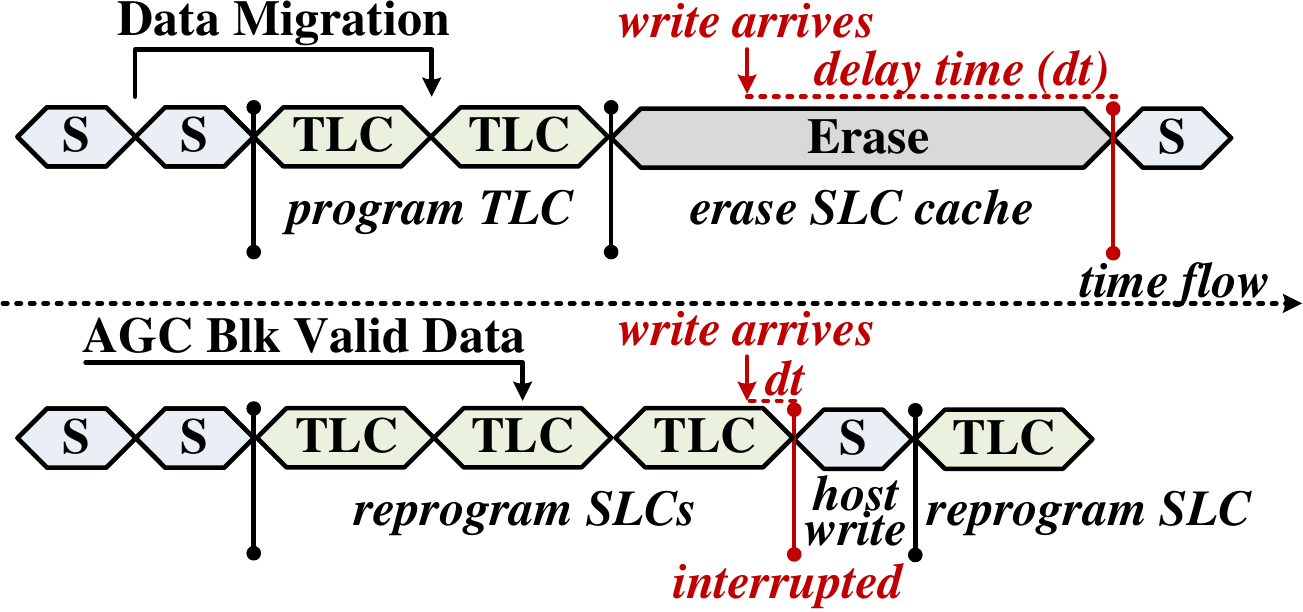}
    \end{minipage}
    \vspace{-0.2in}
\caption{Reprogramming Data from AGC Blocks to SLC cache.}
\label{fig:agc}
    \vspace{-0.1in}
\end{figure}

Take Figure \ref{fig:agc} as an example. 
Assume that the SLC cache is composed of two SLC pages.
For a traditional SLC cache, if a host write arrives suddenly when block reclamation is being performed, it has to be delayed until the reclamation process is finished.
Therefore, the conflict between host write and block reclamation can cause a large delay time.
But for IPS/agc based SLC cache, to deal with the arriving write, valid page migration of AGC is interrupted in time, thus the arriving write can be directed to other SLC pages in the same plane with less delay time.
Note that, reprogram latency is conservatively set to TLC program latency in this work.


\subsection{\textbf{Cooperating IPS/agc and Traditional SLC Cache}}
To meet the demand for a large SLC cache capacity, we propose to cooperate IPS/agc cache with a traditional SLC cache. Due to the limitations of the reprogram operation, IPS/agc can only allocate a small portion of TLC space as an SLC cache. In the proposed cooperative design, the first two layers of the majority of blocks are designated as SLC layers within the IPS/agc cache. Subsequently, the remaining smaller portion of flash blocks is allocated as the traditional SLC cache.

\begin{figure}[!bht]
\vspace{-0.1in}
  \centering
    \begin{minipage}[b]{0.49\textwidth}
      \includegraphics[width=0.98\columnwidth]{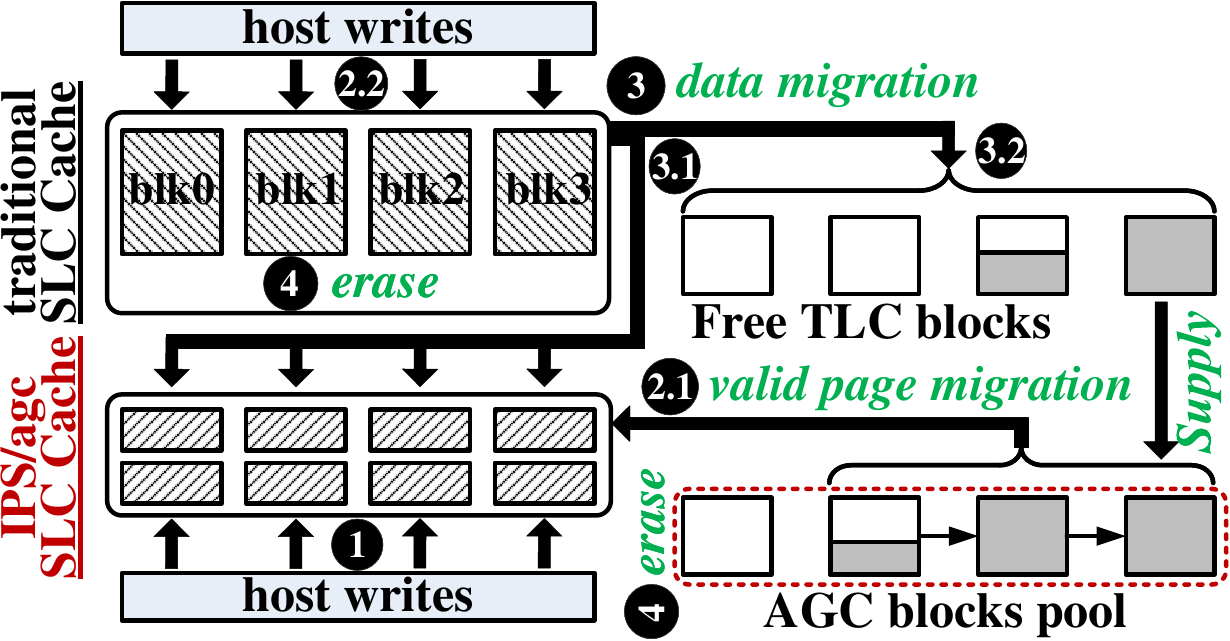}
    \end{minipage}
    \vspace{-0.15in}
\caption{Cooperating IPS/agc and traditional SLC Cache.}
\label{fig:cooperate}
    \vspace{-0.15in}
\end{figure}

In order to relieve performance cliffs and significant write amplification, the IPS/agc cache is prioritized for handling host writes, as shown in Figure \ref{fig:cooperate}. 
First, host writes are directed to the IPS/agc cache (\textbf{Step 1}).
If no additional host writes arrive, valid page migration is performed within the AGC to reprogram the used SLC pages to TLC pages in the IPS/agc cache (\textbf{Step 2.1}). After which a new SLC layer is allocated within the IPS/agc. This cooperative design only activates the traditional SLC cache when sustained host writes require additional space, redirecting subsequent host writes to the traditional SLC cache (\textbf{Step 2.2}).


To ensure that subsequent writes can continuously benefit from the cooperative design, idle time is utilized not only to reprogram used SLC pages in the IPS/agc cache but also to reclaim the traditional SLC cache if it has been consumed.
Since the data migration directions for the IPS/agc cache and the traditional SLC cache are opposite, data from the traditional SLC cache can be read and reprogrammed into the IPS/agc cache. This allows blocks in the traditional SLC cache to be reclaimed while used SLC pages in the IPS/agc cache are reprogrammed (\textbf{Step 3.1}).
During this process, if all SLC pages in the IPS/agc cache have been reprogrammed but there are still blocks in the traditional SLC cache that have not yet been reclaimed, the data in these blocks are redirected to free TLC space (\textbf{Step 3.2}). After completing the data migration, the traditional SLC cache undergoes erase operations (\textbf{step 4}) to reclaim space. Conversely, if all data in the traditional SLC cache has been migrated but the IPS/agc cache still contains used SLC pages that have not been fully reprogrammed, valid page migration from the AGC will fill these gaps.



Thanks to most SLC cache technologies adopt dynamic SLC cache allocation, which determines the size of the SLC cache based on the amount of data stored in SSDs \cite{dynamicwritemicron,inteldy}.
Therefore, in this work, traditional SLC cache in cooperating design can be dynamically allocated without loss of generality.

\subsection{\textbf{Discussion}}
\subsubsection{\textbf{Reliability Issue}}
To ensure reliability, IPS strictly adheres to the reprogram restrictions outlined in \cite{gao2019constructing}:
First, each TLC cell is programmed as SLC, which offers better error tolerance. This means data in the SLC cache can be read and corrected with a lower bit error rate before reprogram operation.
Second, used SLC cells require only two reprogram operations to revert to TLC.
Third, SLC pages are reprogrammed sequentially, minimizing the impact of cell-to-cell interference \cite{cai2017vulnerabilities}.
Quantitatively, each cell in the IPS cache experiences twice the cell-to-cell interference from neighboring cells, while the reliable reprogram method proposed in \cite{gao2019constructing} can tolerate up to twelve times the interference.

\subsubsection{\textbf{Wear Leveling}}
SLC has better endurance than TLC, and wear leveling between them can be achieved by correlating their wear rates. Although Jimenez \textit{et al.} \cite{jimenez2015libra} identified a wear factor between SLC and TLC, it introduces a margin of error that grows with program and erase cycles. This work presents IPS, which achieves wear leveling without depending on a wear factor. Since all flash cells undergo the same wear from program and reprogram operations, IPS uses erase operations as the wear leveling metric. In the cooperative design, wear leveling for the traditional SLC cache is achieved by evenly using TLC space as traditional SLC, while the IPS/agc cache follows the above method.

\subsubsection{\textbf{IPS work with QLC-SLC SSDs}}

Currently, our proposed IPS only operates with TLC-SLC SSDs. Since the voltage margin between the two states of QLC is smaller than that of TLC, the reliability of reprogramming QLC is lower compared to TLC \cite{li2024midas}. Therefore, it is crucial to first study the limitations of reprogramming SLC to QLC. Based on these findings, the SLC cache allocation scheme can be redesigned to minimize the impact of the reprogram operation. In the worst-case scenario, to minimize cell-to-cell interference, reprogram operations may be restricted to within a single layer, thereby reducing the available SLC cache size. However, with the assistance of the proposed cooperating design, IPS can still enhance performance and extend lifespan when host data is prioritized for writing into the SLC cache. 

\section{\textbf{Evaluation}}

\subsection{\textbf{Evaluation Setup}}
\subsubsection{\textbf{Simulated SSDs and Workloads}} In this work, a workload-driven SSD simulator \cite{hu2012exploring} is used, which is configured to match the characteristics of 3D SLC/TLC hybrid SSDs. The evaluation involves a 384GB hybrid SSD, with parameters detailed in Table \ref{tab:ssd}.
Note that, SLC cache is evenly allocated in all planes to exploit the parallelism of SSDs.
The SLC cache size is set to 4GB, which is similar to that of Samsung Turbo Write SSDs \cite{samsungturbo}.
For the proposed cooperative design, the total SLC cache size is increased to 64GB (3.125GB for IPS/agc cache and 60.875GB for traditional SLC cache), roughly matching the size of the real SSD evaluated in Section \ref{sec:moti}.   
The workloads studied in this work include a subset of MSR Cambridge Workloads from servers \cite{Narayanan2009,gao2019constructing}.

\renewcommand\arraystretch{0.95}
\begin{table}[!htb]
\vspace{-0.1in}
\footnotesize
\begin{center}
\caption{Parameters of the Simulated SSD \cite{gao2019constructing}}
\begin{tabular}{lllll}
\cline{1-2}
\multicolumn{1}{|l||}{\textbf{\begin{tabular}[c]{@{}l@{}}SSD \\ Configuration\end{tabular}}} & \multicolumn{1}{l|}{\begin{tabular}[c]{@{}l@{}}384GB; 8 Channels; 4 Chips/Channel; 2 Die/Chip; \\2 Planes/Die; 2048 Blocks/Plane; 384 Pages/Block; \\4KB Page.\end{tabular}} &  &  &  \\ \cline{1-2}
\multicolumn{1}{|l||}{\textbf{\begin{tabular}[c]{@{}l@{}}Timing \\ Parameters\end{tabular}}} & \multicolumn{1}{l|}{\begin{tabular}[c]{@{}l@{}}0.02 ms for SLC read; 0.066 ms for TLC read; \\ 0.5 ms for SLC write; 3 ms for TLC write;\\ 10 ms for block erase.\end{tabular}}            &  &  &  \\ \cline{1-2} &  &  &  &  \\
            &  &  &  &
\end{tabular}
\label{tab:ssd}
\end{center}
\vspace{-0.35in}
\end{table}

\subsubsection{\textbf{Evaluated Schemes}}
The baseline in this work adopts Turbo Write technology \cite{samsungturbo}.
First, the SLC cache size is set to 4GB for baseline, IPS (without AGC assistance), and IPS/agc (with AGC assistance).
Then, the cooperating design is implemented with a 64GB SLC cache size.
Although dynamic SLC cache allocation is supported in most modern hybrid SSDs, it is realized before processing host writes.
Therefore, the size of the SLC cache is fixed in this evaluation for simplicity.

\subsection{\textbf{Results and Analysis}}
\subsubsection{\textbf{Results of IPS}}
\begin{figure}[!htb]
 \vspace{-0.05in}
  \subfloat[Bursty Access.]{
    \label{fig:latency-b}
    \begin{minipage}[b]{0.237\textwidth}
      \includegraphics[width=0.99\columnwidth]{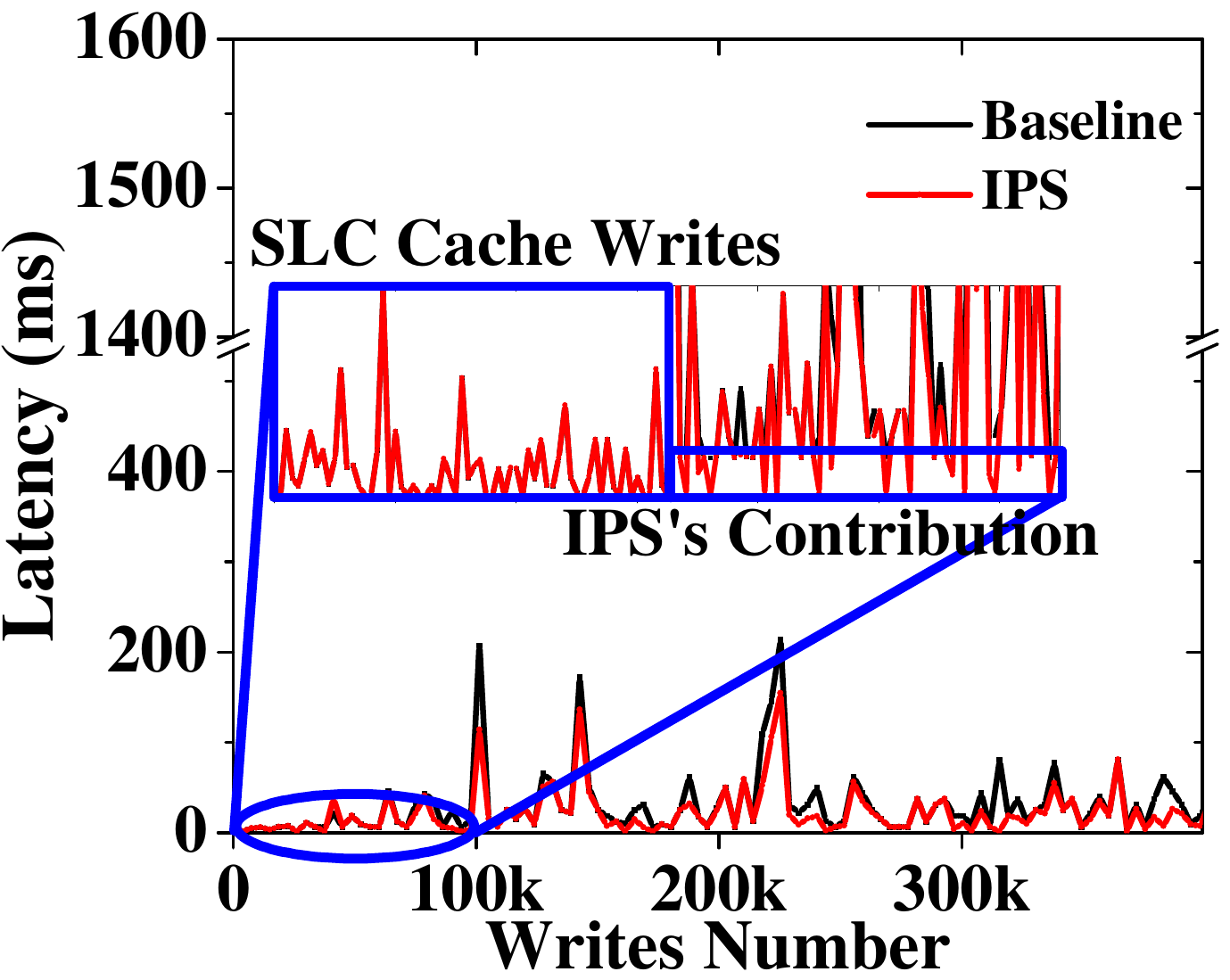}
      \vspace{-0.2in}
    \end{minipage}}
  \subfloat[Daily Use.]{
    \label{fig:latency-d}
    \begin{minipage}[b]{0.229\textwidth}
      \includegraphics[width=0.99\columnwidth]{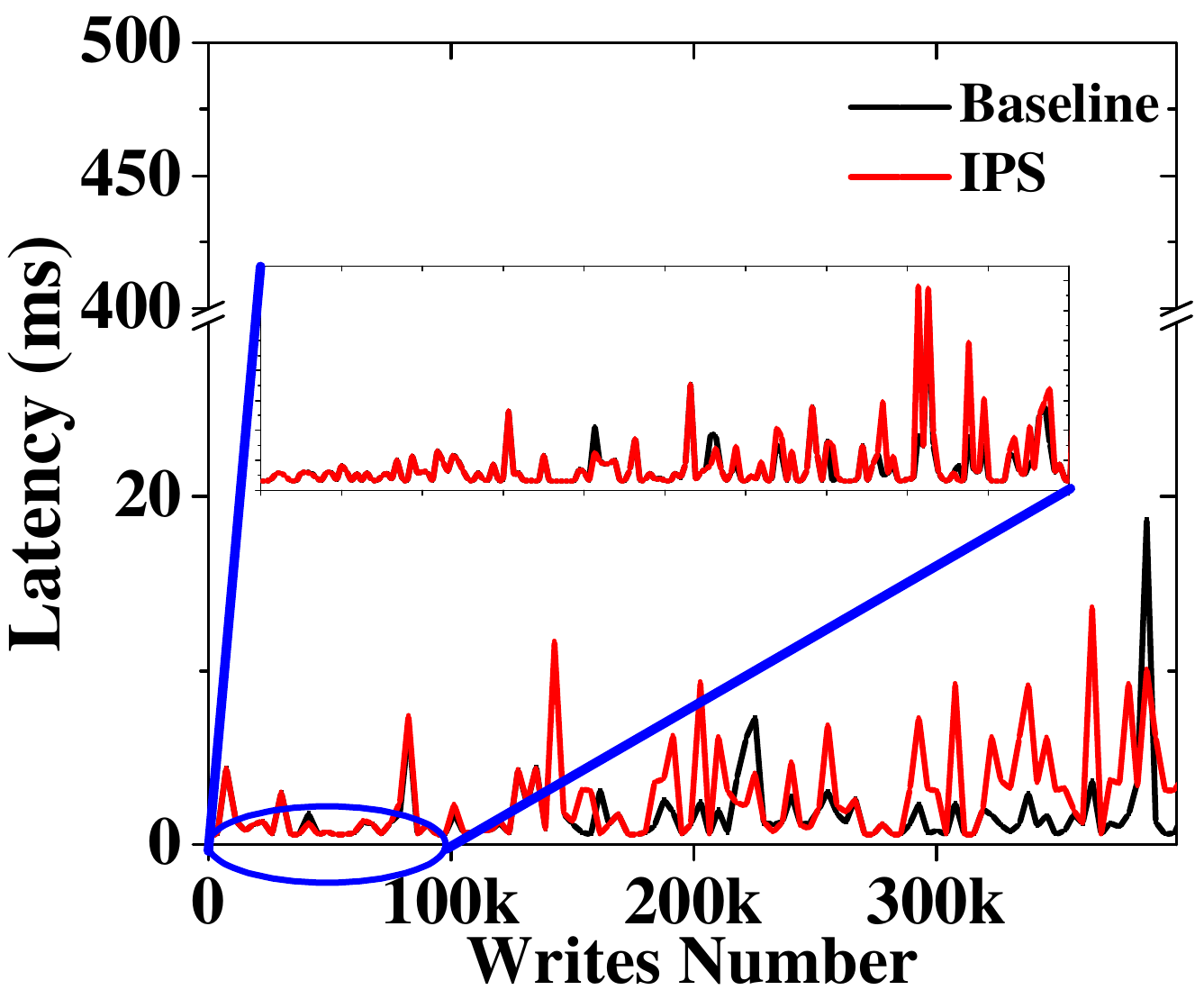}
      \vspace{-0.2in}
    \end{minipage}}
    \vspace{-0.05in}
\caption{Write Latencies during Runtime.}
\vspace{-0.13in}
\label{fig:latency}
\end{figure}

First, for bursty access, Figure \ref{fig:latency-b} shows the write latencies of baseline and IPS during runtime.
Take HM\_0 as an example.
For the first 100,000 writes, we can have three observations:
First, in the left blue box, IPS achieves the same write latency with baseline because all writes are programmed at the SLC performance level;
Second, when the SLC cache is filled up, the latencies of both schemes are significantly increased to the TLC performance level;
Third, in the right blue box, compared with the baseline, IPS can achieve lower write latency as the new SLC cache is intermittently allocated.

\begin{figure}[!htb]
  \vspace{-0.05in}
  \subfloat[Bursty Access.]{
    \label{fig:exp-b}
    \begin{minipage}[b]{0.235\textwidth}
      \includegraphics[width=0.99\columnwidth]{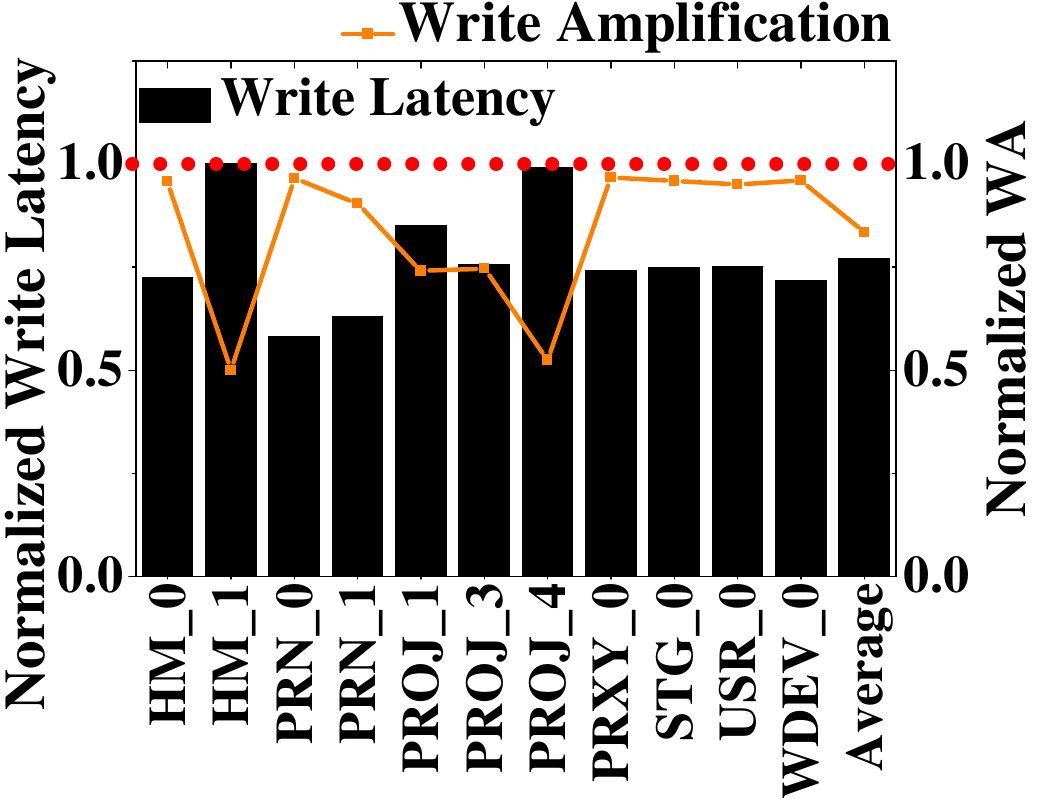}
      \vspace{-0.2in}
    \end{minipage}}
  \subfloat[Daily Use.]{
    \label{fig:exp-d}
    \begin{minipage}[b]{0.235\textwidth}
      \includegraphics[width=0.99\columnwidth]{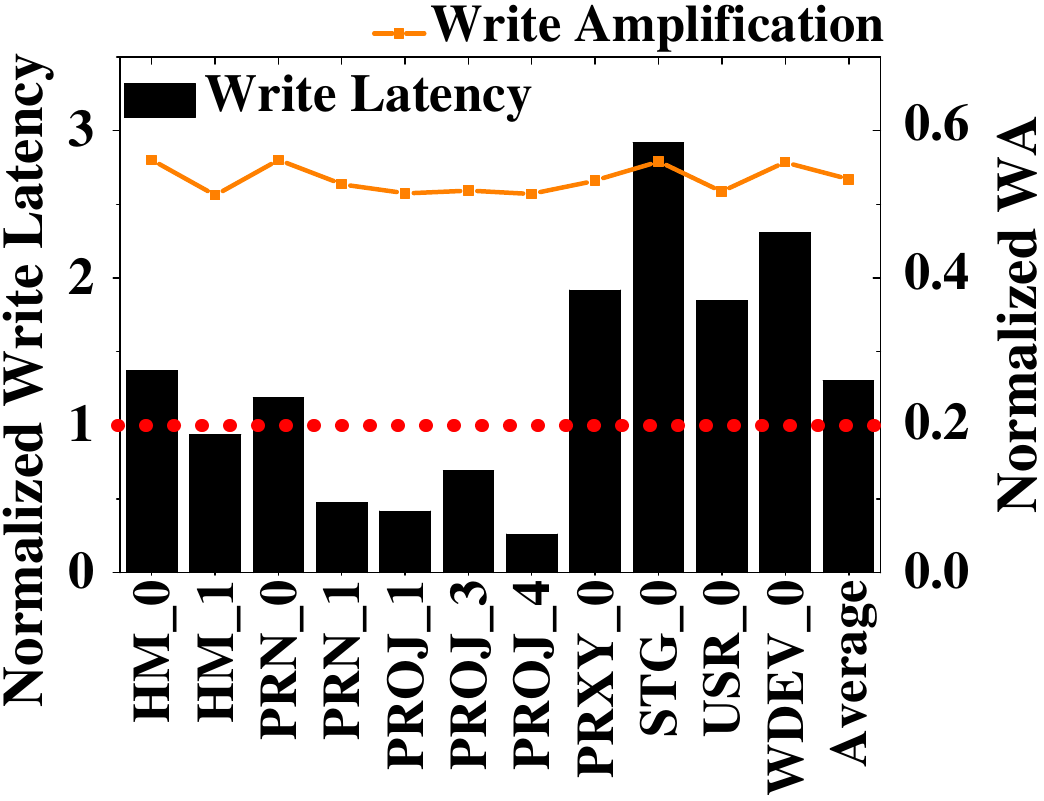}
      \vspace{-0.2in}
    \end{minipage}}
    \vspace{-0.05in}
\caption{Normalized Write Latency and Write Amplifications.}
\vspace{-0.2in}
\label{fig:exp}
\end{figure}

In Figure \ref{fig:exp-b}, the average write latency of IPS is collected and normalized to baseline.
On average, IPS reduces write latency by 0.77 times compared to the baseline. However, for HM\_1 and PROJ\_4, IPS achieves similar write latency to the baseline because most data is programmed at SLC performance levels. In terms of write amplification, IPS reduces it by nearly half for these workloads, as all SLC cache data must be migrated to TLC space. For other workloads, where the SLC cache is quickly depleted and most data is written directly to TLC space, the reduction in write amplification is modest. Overall, IPS reduces write amplification by 0.83 times compared to the baseline.

For daily use, evaluated results of baseline and IPS are collected and presented in Figure \ref{fig:exp-d}.
Similarly, for the first 100,000 writes of HM\_0, we have two observations:
First, the write latency is gradually increased due to the conflict between block reclamation and host writes.
Second, compared with the baseline, the write latency of IPS is larger, which is increased by 1.3 times, as shown in Figure \ref{fig:exp-d}.
However, there also are several workloads that achieve lower write latency over baseline, such as PROJ\_4.
Due to its small total write size, no reprogram operations are operated for PROJ\_4.
But for baseline, host writes still suffer from the conflict between time-consuming block reclamation and host writes, causing write latency to increase.
For write amplification, on average, IPS reduces it by 0.53 times compared to the baseline.
This is because only a small amount of data is updated before data migration.
Therefore, IPS which does not cause write amplification can almost reduce write amplification to half.

\subsubsection{\textbf{Results of IPS/agc}}
\begin{figure}[!bht]
    \vspace{-0.05in}
  \centering
    \begin{minipage}[b]{0.41\textwidth}
      \includegraphics[width=0.98\columnwidth]{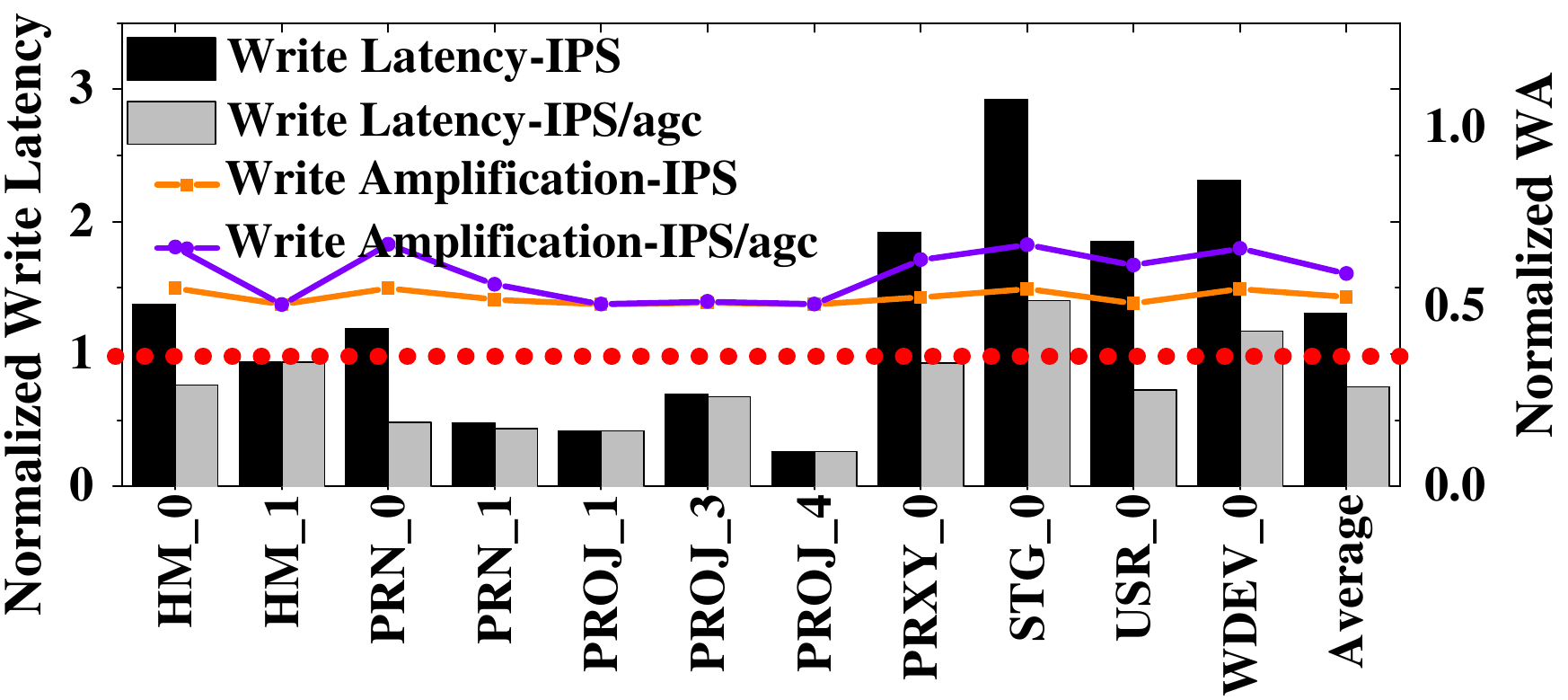}
    \end{minipage}
    \vspace{-0.1in}
\caption{Normalized Results of IPS and IPS/agc.}
\label{fig:res agc}
    \vspace{-0.2in}
\end{figure}

In this section, write latency and write amplification of IPS and IPS/agc are normalized to baseline and presented in Figure \ref{fig:res agc}.

For write latency, compared with baseline, IPS/agc can averagely reduce it by 0.75 times while IPS increases write latency by 1.3 times.
But there are two exceptions for IPS/agc (STG\_0 and WDEV\_0), of which write latencies still are larger than that of baseline.
The main reason is that IPS/agc still can suffer from reprogram operations as used SLC pages have not been fully reprogrammed when host writes arrive.

For write amplification, AGC may increase it when valid pages that will be invalidated soon are migrated in advance.
During the evaluation, write amplification resulting from AGC is counted into IPS/agc.
Thus, IPS/agc averagely increases write amplification by 0.07 times compared to IPS.
Compared with baseline, IPS/agc can reduce write amplification by 0.59 times, on average.

\subsubsection{\textbf{Results of cooperating Design}}
\begin{figure}[!htb]
  \vspace{-0.1in}
  \subfloat[Bursty Access.]{
    \label{fig:co-b}
    \begin{minipage}[b]{0.23\textwidth}
      \includegraphics[width=0.99\columnwidth]{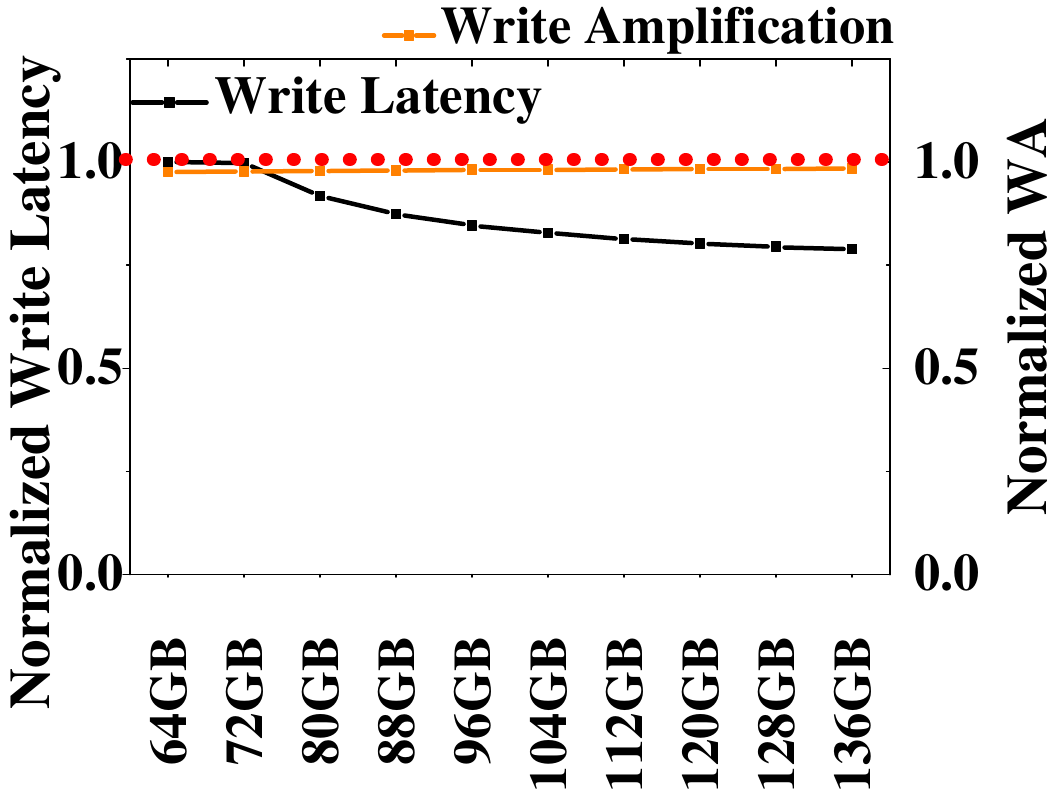}
      \vspace{-0.2in}
    \end{minipage}}
  \subfloat[Daily Use.]{
    \label{fig:co-d}
    \begin{minipage}[b]{0.23\textwidth}
      \includegraphics[width=0.99\columnwidth]{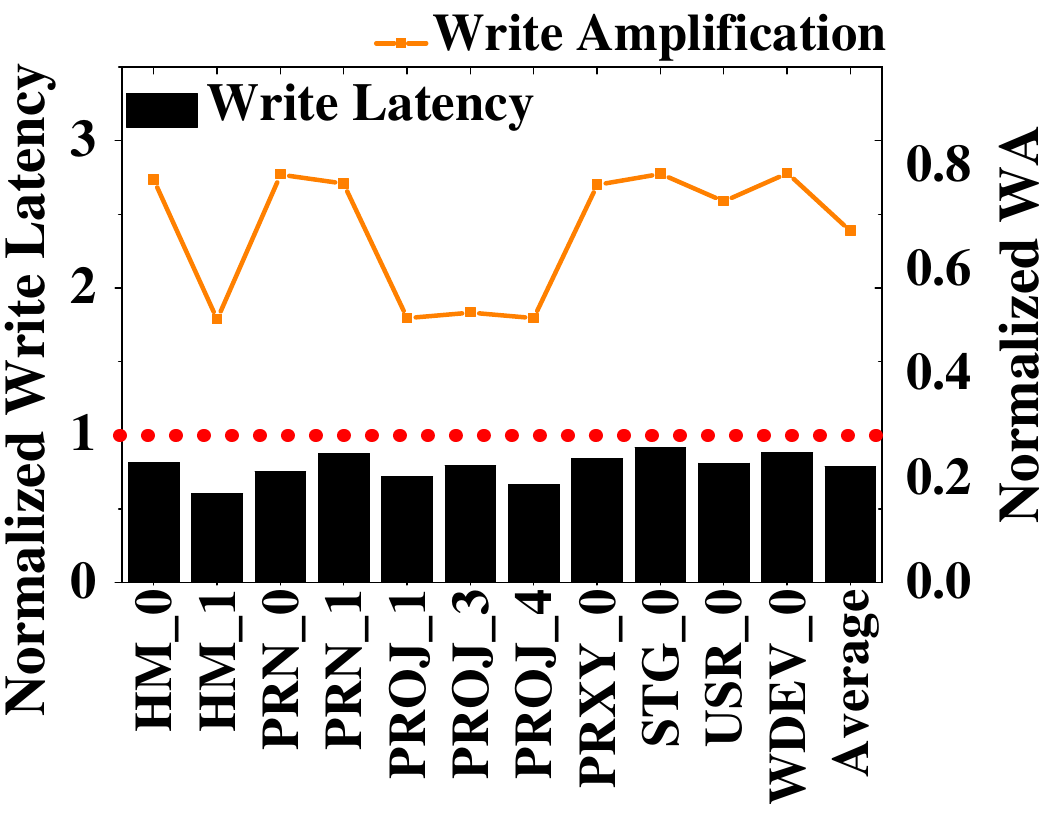}
      \vspace{-0.2in}
    \end{minipage}}
    \vspace{-0.05in}
\caption{Normalized Write Latency and Write Amplifications.}
\vspace{-0.1in}
\label{fig:co}
\end{figure}

In this section, write latency and write amplification of cooperating design are evaluated and normalized to baseline.
For bursty access, take HM\_0 as an example.
Total write size is varied from 64GB to 136GB by running workload repeatedly.
As shown in Figure \ref{fig:co-b}, when the total write size is 64GB, all data can be written into the SLC cache, and the cooperating design has the same write latency as that of the baseline.
Similarly, cooperating design only can achieve a slight write amplification decrease as only less data is written into the IPS/agc cache.
However, with the increase in total write size, more writes can benefit from new allocated IPS/agc cache so that the write latency of the cooperating design is reduced.
When the total write size is 136GB, the cooperating design can averagely reduce write latency by 0.79 times compared to baseline.
On the contrary, normalized write amplification is gradually approaching 1 with the increase of total write size.
This is because baseline programs more writes into TLC space without causing write amplification.
Therefore, the contribution from IPS/agc gradually dwindles.
When the total write size is 136GB, the cooperating design only can reduce write amplification by 0.98 times compared to the baseline.

For daily use, workloads are executed repeatedly to increase the total write size.
In this evaluation, we set the total write size to 64GB as results with larger write sizes are similar.
Results of cooperating design are normalized to baseline and presented in Figure \ref{fig:co-d}.
The cooperating design achieves similar results to the baseline for some workloads, such as STG\_0, due to sufficient idle time for data migration, which allows most data to be written to the SLC cache. In contrast, for workloads like PROJ\_4, the cooperating design significantly reduces write latency compared to the baseline. This improvement is attributed to the higher priority given to IPS/agc writes, which allows host writes to be delayed with reduced time costs. On average, the cooperating design decreases write latency by 0.78 times and reduces write amplification by 0.67 times compared to the baseline.

\section{\textbf{Conclusion}}
In this work, firstly, we evaluate emerging hybrid 3D SSDs and deliver the potential pitfalls that cause performance decreases and significant write amplification.
Then, reprogram operation is used to conduct a new SLC cache without generating additional writes, thus data migration of the traditional SLC cache is removed from the critical path.
As a result, evaluated results show that the proposed IPS can effectively improve write performance and reduce write amplification.

\bibliographystyle{plain}
\bibliography{references}

\end{document}